# CONSTRUCCIÓN DE UN REACTOR DE NITRURACIÓN POR PLASMA Y SIMULACIÓN DE LA DISTRIBUCIÓN DE TEMPERATURAS

**Javier García Molleja[a], Pablo Romero-Gómez[b] y Jorge N. Feugeas[a]**

[a]*Instituto de Física Rosario (CONICET – Universidad Nacional de Rosario), Bvrd. 27 de febrero 210 bis, S2000EZP, Rosario, Argentina*
[b]*ICFO – Institut de Ciències Fotòniques, Parque Mediterráneo de la Tecnología, E-08860, Castelldefels, España*
*E-mail (J. García Molleja): garciamolleja@ifir-conicet.gov.ar*

**Palabras claves:** Aceros, Plasma, Reactores de Nitruración, Simulación Numérica, Temperatura

**RESUMEN**
*El tratamiento de aceros austeníticos mediante nitruración y cementación por plasma permite mejorar las propiedades mecánicas de éstos, tales como su dureza y resistencia al desgaste, sin alterar sus buenas propiedades anticorrosivas. Para una perfecta caracterización y reproducibilidad del proceso es necesario tener en cuenta, además de los parámetros de trabajo, las variables ocultas del sistema. En este trabajo se indica el diseño de un cátodo especial (colocado en un nuevo reactor) que permite el encastre de probetas para eliminar el efecto de borde. Tras la construcción y demostración del buen funcionamiento, se simula numéricamente la distribución de temperaturas en el interior de la cámara creada por las colisiones de moléculas de nitrógeno con el cátodo, determinando que la refrigeración externa es un aspecto vital a considerar. La resolución se consigue a partir de la ecuación de Laplace con el método de Jacobi paralelizado. Mediante espectroscopia óptica de emisión se determina el mecanismo de transferencia de calor mediante conducción y se verifican experimentalmente los resultados obtenidos con la simulación.*

## 1. INTRODUCCIÓN

La cementación [1] y la nitruración [2] por plasma son las técnicas actuales que lideran el tratamiento superficial de aceros inoxidables aplicados a multitud de campos industriales y de ingeniería. La modificación superficial de los aceros inoxidables, especialmente los austeníticos (AISI 316L, por ejemplo [3]), es un objetivo primordial debido al aumento de dureza y a la gran resistencia al desgaste que otorgan estas técnicas por la creación de la llamada austenita expandida, todo ello sin mermar la buena resistencia a la corrosión que poseen naturalmente este tipo de aceros [4]. También es digno de considerar los bajos costos que tienen dichas técnicas, así como su facilidad de repetición y la ausencia de contaminantes.
Sin embargo, la mayoría de estudios llevados a cabo sólo atienden a vincular la realización experimental con los parámetros de control del plasma (tensión, presión, mezcla de gases, temperatura) dejando de un lado la forma y configuración del reactor. Por consiguiente, y este trabajo atenderá al diseño y construcción de un cátodo que se empleará en los experimentos. Así como una simulación numérica encargada de determinar cómo se distribuye la temperatura en su interior, para conocer cuán fundamental es la refrigeración externa y dónde es propicio colocar las probetas para un tratamiento uniforme.

## 2. CONSTRUCCIÓN DEL DISPOSITIVO

Tras construir el reactor con acero AISI 304 y tomar precauciones para evitar fugas reales y virtuales [5], es necesario diseñar un cátodo adecuado y que esté bien aislado de la cámara (que actuará como ánodo), esquematizada en la parte izquierda de la Figura 1. Los efectos de borde son un fenómeno crucial para el tratamiento superficial, ya que el aumento de campo eléctrico en éstos hace que no toda la superficie se nitrure de igual modo. Existen varias configuraciones, tales como la caja catódica [6], o encastrar las probetas dentro del cátodo para que el efecto de borde se dé en otro lugar. En el primer caso, por su incipiente estado de estudio, no se alcanzaron resultados idénticos a la nitruración clásica, existiendo incluso contaminantes de la caja [7]. Para el segundo, si bien está utilizado ampliamente, no hay indicaciones de qué materiales son ideales (diferencias de coeficientes de dilatación térmica, resistividad eléctrica…) o el tamaño de los agujeros donde van las probetas.

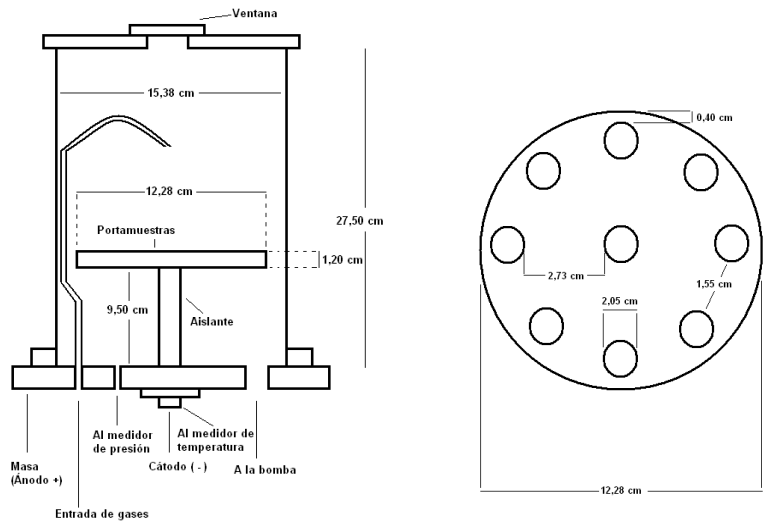

**Figura 1. Izquierda:** Esquema del reactor de nitruración. **Derecha:** Diseño del cátodo y medidas.

Nuestro diseño, esquematizado en la parte de la derecha de la Figura 1, demuestra que el acero AISI 304 puede usarse en la construcción, puesto que sus propiedades físicas, mecánicas y térmicas [8] son muy semejantes a las del acero AISI 316, que es el más utilizado en nuestros tratamientos. Con un gran diámetro se podrán colocar multitud de probetas, maximizando el proceso para cada sesión. En nuestro caso, el cátodo consta de una placa anclada en la alimentación (donde se colocan las probetas) y otra de espesor igual a las probetas y con agujeros para que el sistema quede todo al mismo nivel. El diámetro de los agujeros es un poco mayor (0,05 cm) que el diámetro de las probetas para lograr una fácil colocación de éstas (un diámetro menor dificultaría colocar correctamente la probeta) sin tener propensión a la formación del fenómeno denominado cátodo hueco, que elevaría drásticamente la temperatura.

## 3. SIMULACIÓN DE LA DISTRIBUCIÓN DE TEMPERATURAS

3.1 Planteamiento del problema y discretización

Es esencial determinar cómo se distribuirá la temperatura dentro de la cámara, puesto que un alto gradiente de temperaturas afectará la difusión del nitrógeno dentro de la estructura del acero austenítico. Como punto de partida se indica que la cámara posee una refrigeración externa mediante una serpentina en la que circula agua. Ésta se localiza en la parte inferior del cilindro y se utiliza para evitar la rápida degradación de los o-rings que sellan el vacío. Mediante una simulación numérica podemos ver si con esta configuración es suficiente para poder generar gradientes de temperatura poco acusados o se hace necesario cubrir toda la altura de la cámara con la serpentina.

La resolución de este problema se lleva a cabo mediante la aplicación de la *ecuación de Poisson*, en la que identificamos al campo escalar *u(x,y)* como la temperatura que se origina mediante la transferencia de momento de los iones de nitrógeno molecular que bombardean el cátodo al llegar a él acelerados por la tensión eléctrica impuesta.

$$\frac{\partial^2 u(x,y)}{\partial x^2} + \frac{\partial^2 u(x,y)}{\partial y^2} = f(x,y) \quad (1)$$

Esta Ecuación 1 describe un problema físico que está bien condicionado (respecto su planteamiento matemático) en cuanto a su aplicabilidad en tratamientos numéricos. Si el término *f(x,y)*, que representa la existencia de fuentes y sumideros, tiene un valor de 0, estamos en el caso particular de la ecuación de Poisson, denominada de *Laplace*, indicada en la Ecuación 2:

$$\frac{\partial^2 u(x,y)}{\partial x^2} + \frac{\partial^2 u(x,y)}{\partial y^2} = 0 \qquad (2)$$

Las Ecuaciones 1 y 2 se aplican a un dominio bidimensional, puesto que conseguiremos así un ahorro computacional importante. De todas maneras, la solución armónica que obtengamos, y representemos como un conjunto de curvas de nivel, es idéntica si tomamos para simular otro plano de corte perpendicular al primero, debido a la simetría cilíndrica de la cámara. Debido al poco uso industrial de este tipo de tratamiento (los basados en calentamiento por corriente catódica sin fuentes de calor auxiliares) frente a los reactores de calefacción auxiliar, esta expresión de la Ecuación 2 apenas se ha aplicado a la caracterización de reactores de nitruración. Por tanto, resolviéndola se tendrá conocimiento de la distribución de temperaturas. Este dato, junto con otras variables ocultas (entrada de gases, velocidad de bombeo), es necesario para poder discernir el por qué de tantos resultados divergentes para iguales condiciones experimentales (presión de trabajo, flujo de gases, temperatura, tensión aplicada, etc.).

En realidad, el cátodo va a actuar como la fuente de temperatura, mientras que las paredes de la cámara serán los sumideros, pero se puede hacer la argucia matemática [9] de suponer que dichas zonas son contornos, por lo que fijando las condiciones se obtendrá el mismo resultado sin la complicación añadida de manejar el término *f(x,y)*.

Las mencionadas condiciones de contorno se miden experimentalmente mediante un pirómetro de reflectancia OAKTRON de potencia < 1 mW, 95% de eficiencia y longitud de onda entre 630 y 670 nm. Dicha medición se realizó tras seguir los pasos típicos de un proceso de nitruración: la cámara alcanzó un vacío residual de 20 mTorr, llenándose posteriormente de nitrógeno molecular de alta pureza hasta alcanzar una presión de 3,750 Torr. La descarga se mantuvo en valores de tensión y corriente típicos de una sesión de nitruración (700 V y 450 mA, lo que implica una densidad de corriente de 1,70 mA/cm$^2$) para que el cátodo llegase a una temperatura de 400 ºC. Las paredes alcanzaron valor estacionario de temperatura a los 80 minutos de llegar a la condición de medición. En la Figura 3 quedan detalladas las condiciones de contorno.

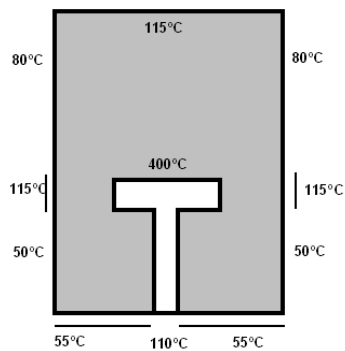

**Figura 3.** Diagrama de los valores del contorno para la simulación.

La resolución de la ecuación de Laplace con unas condiciones de contorno impuestas se denomina *problema de Dirichlet*. Una vez conocidos estos datos es necesaria una discretización del problema para poder resolverlo aplicando el método de *Jacobi* paralelo. La discretización del problema para implementar simulación conlleva que todos los puntos de contorno queden considerados en un mallado en el que el tamaño de paso es idéntico en ambas dimensiones. Como las el valor de altura es diferente al de anchura (la mitad de este valor) el número de puntos va a ser diferente, tal y como se deduce de la Expresión 3:

$$h = \frac{L_x}{N_x + 1} = \frac{L_y}{N_y + 1} \qquad (3)$$

La relación entre el número de puntos entre una dimensión y otra queda clara. Por facilidad de asignación de fronteras se elige que ambos valores sean múltiplos de 32. El valor de 400 ºC para el cátodo queda fijo utilizando una máscara para que las iteraciones de la simulación no afecten a dicha región. No se asigna una máscara al aislante para poder determinar así qué gradiente de temperaturas soporta. En cuanto a las condiciones de contorno, serán considerados como nodos *frontera* y no se actualizarán. La discretización de las derivadas segundas parciales se consigue aplicando la fórmula de los *cinco puntos*, detallada en la Ecuación 4 y adaptada para el método de Jacobi, en la que se conoce el valor del punto $U(x_i,y_j)$ en la iteración *k+1* a partir de los valores adyacentes en la iteración *k* [10].

$$U_{i,j}^{k+1} = \frac{1}{4}[U_{i+1,j}^{k} + U_{i-1,j}^{k} + U_{i,j+1}^{k} + U_{i,j-1}^{k}] \qquad (4)$$

3.2 Resultados de la simulación

La simulación se hará utilizando el programa FORTRAN90 y recurriendo a las directivas !HPF$, puesto que el método de Jacobi puede ser paralelizado, distribuyendo por bloques la matriz de puntos entre procesadores. Los cálculos se llevan a cabo con cuatro procesadores doble Intel Xeon a 3,2 GHz y 2Gb de RAM del clúster *Beowulf* del Grupo de Simulación Numérica de la Universidad de Córdoba (España). La tolerancia adjudicada es de 0,00001 y el número máximo de iteraciones de control será 100000. En la Figura 4 se detalla la solución obtenida tras realizar la simulación tanto en curvas de nivel como en superficies. Experimentos y simulaciones idénticos con el cátodo a 100, 200 y 300 ºC dan distribuciones de temperatura bastante similares, dando cuenta que la estructura y refrigeración son los que determinan la forma de la solución.

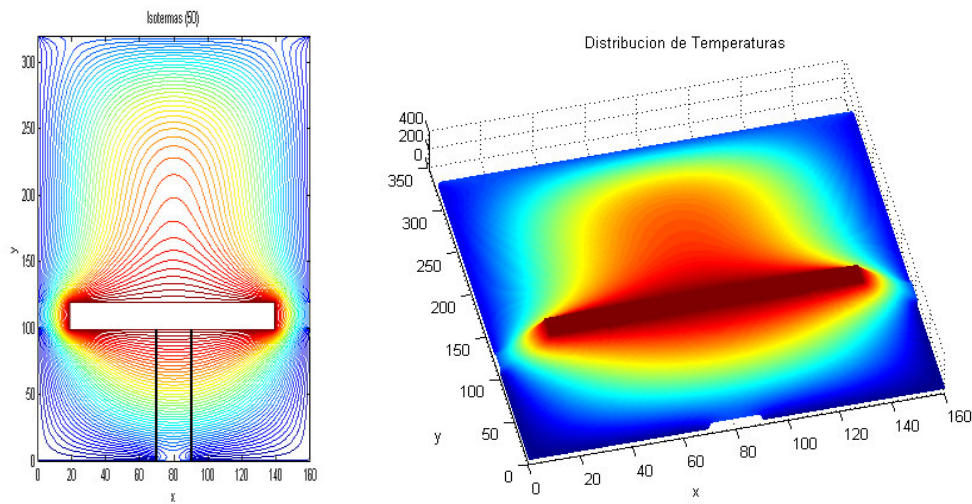

**Figura 4. Izquierda:** Solución mediante curvas de nivel. **Derecha:** Distribución de temperaturas mediante superficies (Rojo: alta temperatura, Azul: baja temperatura).

Se observa que por encima del cátodo el gradiente es bajo, lo que resulta un dato óptimo a la hora de nitrurar. Sin embargo, en los bordes es más acusado, siendo entonces necesario situar las probetas en una zona no tan alejada del centro. La parte inferior muestra un alto gradiente, por lo que se hace necesario una mayor refrigeración para preservar la integridad de los aislantes. La conclusión entonces es el requerimiento de ampliar el recorrido de la serpentina por toda la altura de la cámara.

El uso de cuatro procesadores fue útil para lograr un ahorro importante de tiempo de ejecución. Si nuestro dominio cuenta con 540800 puntos podemos repetir la simulación en modo secuencial y en paralelo con 1, 2 y 4 procesadores. El tiempo de ejecución variará. Para ver la mejora de uno respecto a otro podemos dividir el tiempo de ejecución secuencial por el tiempo en paralelo para obtener la ganancia en velocidad $S_p$. Si este valor se divide por el número de procesadores se obtiene el valor de la eficiencia η. Esto queda reflejado en la Tabla 1.

**Tabla 1**. Valores del tiempo de ejecución, de ganancia en velocidad y eficiencia para la ejecución en secuencial y paralelo con 1, 2 y 4 procesadores

| Procesadores | T (s) | $S_p$ | η |
|---|---|---|---|
| **1 (secuencial)** | 11,5143 | - | - |
| **1 (paralelo)** | 11,8920 | 0,968239 | 0,96823 |
| **2** | 6,28200 | 1,832904 | 0,91645 |
| **4** | 3,64500 | 3,158930 | 0,78973 |

## 4. MEDICIÓN DE TEMPERATURA DE GAS POR ESPECTROSCOPIA ÓPTICA DE EMISIÓN

Se utiliza para captar la radiación procedente de la descarga un monocromador Jarrell-Ash con un montaje Czerney-Turner de longitud focal de 275 mm. La rendija de entrada se impuso a 25 μm, mientras que la red de difracción empleada constaba de 3600 líneas/mm. La luz se focaliza por una lente antes de llegar al monocromador y como detector se emplea un arreglo lineal de 1024 fotodiodos. Por consiguiente, la resolución del dispositivo es de aproximadamente de 0,05 nm. El análisis que se lleva a cabo barre longitudes de onda situadas entre 300 y 500 nm. Por limitaciones evidentes en la construcción del reactor de nitruración, las mediciones con OES se llevaron a cabo en otra cámara de iguales dimensiones (con el mismo cátodo de encastre que empleamos en la cámara diseñada), pero que posee una ventana lateral.

Debido a la cinética del brillo negativo presente en estos tipos de descarga (DC a baja presión), los mecanismos de excitación para las bandas del segundo sistema positivo ($N_2(C) \rightarrow N_2(B) + h\nu$) en descargas de $N_2$ obedecen a la Ecuación 5 [11]:

$$e + N_2(X,0) \Rightarrow e + N_2(C,v') \qquad (5)$$

Además, el nivel $N_2^+(B)$ queda poblado principalmente por colisiones de electrones con el estado fundamental de $N_2$, es decir, como indica la Ecuación 6:

$$N_2(X,0) + e \Rightarrow N_2^+(B,v') + e + e \qquad (6)$$

En ambos casos, el proceso principal de desexcitación es el decaimiento radiativo.

Cuando una molécula colisiona con un electrón es esperable que no sea alterado el momento rotacional. Por tanto, se puede asumir que la distribución rotacional no cambia en la excitación mediante colisión de electrones [11]. Es decir, si la excitación se produce por los procesos 1 y 2, la distribución de los niveles $N_2(C)$ y $N_2^+(B)$ corresponderán a la distribución rotacional del nivel fundamental $N_2(X^1\Sigma_g^+)$.

Todo esto muestra que la temperatura del gas se puede determinar a partir de la distribución rotacional de las bandas del segundo sistema positivo ($N_2(C) \rightarrow N_2(B) + h\nu$) y también del primer sistema negativo ($N_2^+(B) \rightarrow N_2^+(X)$).

La técnica de OES también sirve para confirmar que existe una tendencia idéntica del gradiente de temperaturas con los datos obtenidos por simulación. Hay que tomar en cuenta que el modelo teórico que traduce intensidad de la radiación emitida por el plasma en temperatura puede sobreestimar los valores [12], por lo que la comparación sólo puede ser cualitativa. La Figura 5 representa la medición experimental de la caída de temperatura por encima del cátodo y un corte perpendicular de valores de la distribución de temperaturas obtenidas por simulación.

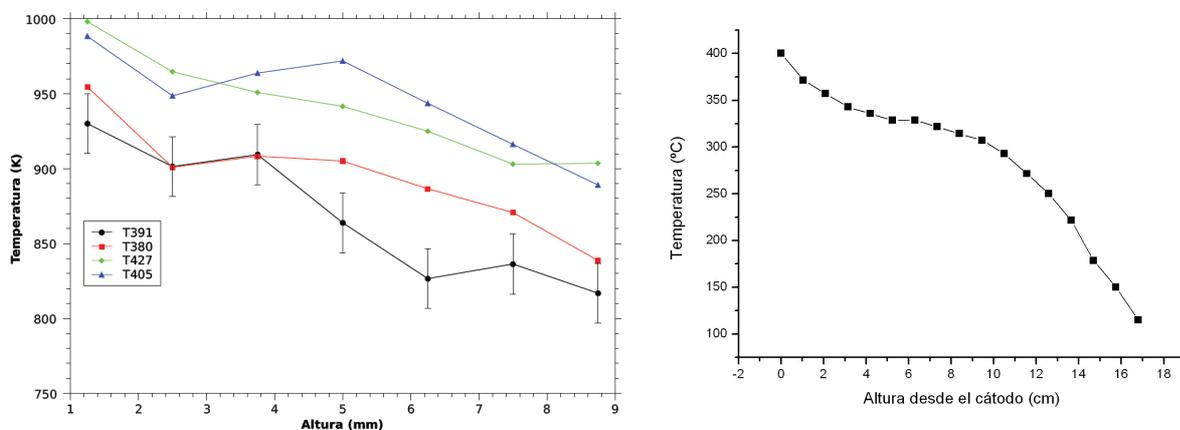

**Figura 5. Izquierda:** Valores de temperatura sobre el cátodo de las especies $N_2$ (427 y 405) y $N_2^+$ (391 y 380) medidas en dos bandas. **Derecha:** Corte transversal de la distribución de temperaturas de la simulación.

De todas maneras, existe en los resultados experimentales una caída de temperatura más acusada que en la simulación a grandes distancias, por lo que para mejorar el modelo hay que tener en cuenta los procesos de convección juntos a los de la conducción (ya que la radiación queda solapada a estas temperaturas).

**5. CONCLUSIONES**

La descripción de la construcción del reactor es necesaria para poder dar a conocer las variables ocultas que se dan en nuestro equipo, al no existir un consenso internacional de diseño estandarizado, dando ayuda a grupos interesados en la nitruración. El diseño del cátodo también ayuda a aclarar los motivos por los que el encastre se está utilizando cada vez más en los experimentos.

La simulación numérica determina que por arriba del cátodo el gradiente es bajo y que en los bordes es más acusado, por lo que las probetas han de quedar colocadas en una posición no muy excéntrica. Esta conclusión se verifica por hechos experimentales donde se ha observado mediante difracción de rayos X de algunas probetas de acero picos pertenecientes a precipitados de nitruros de Cr y Fe. La refrigeración de nuestro equipo se demuestra insuficiente, siendo por tanto necesario cubrir toda la altura de la cámara. Los experimentos apoyan estas conclusiones, además de haberse llevado a cabo una determinación del mecanismo de transmisión del calor.

La simulación en paralelo determina las mejoras de tiempo que se obtienen para realizar cálculos laboriosos y se demuestra que con cuatro procesadores la eficiencia es mayor a pesar de una mayor demora por tiempo de comunicación entre procesadores.